\documentclass[10pt,conference]{IEEEtran}

\usepackage{amsmath}
\usepackage{amsfonts}
\usepackage{amssymb}
\usepackage{times}
\usepackage{psfrag}
\usepackage{cite}
\usepackage{stfloats}
\usepackage{amsmath}
\usepackage{enumerate}
\usepackage{amsfonts}
\usepackage{amssymb}
\usepackage{color}
\usepackage{psfrag}
\usepackage{cite}
\usepackage{subfigure}
\usepackage{epsfig}
\newtheorem{theorem}{Theorem}
\newtheorem{lemma}{Lemma}

\newtheorem{proposition}{Proposition}

\makeatletter
\newcommand{\smatlabaxislabel}[1]{\fontsize{12}{\f@baselineskip}%
\textsf{#1}}
\newcommand{\matlabaxislabel}[1]{\fontsize{14.4}{\f@baselineskip}%
\textsf{#1}}
\newcommand{\mmatlabaxislabel}[1]{\fontsize{17.28}{\f@baselineskip}%
\textsf{#1}}
\newcommand{\bmatlabaxislabel}[1]{\fontsize{20.74}{\f@baselineskip}%
\textsf{#1}}
\newcommand{\bbmatlabaxislabel}[1]{\fontsize{24.88}{\f@baselineskip}%
\textsf{#1}} \makeatother

\allowdisplaybreaks

\centerfigcaptionstrue

\begin{document}

\title{Full Diversity Blind Signal Designs for Unique Identification of Frequency Selective Channels}

\author{\authorblockN{Jian-Kang Zhang}
\authorblockA{Department of Electrical and Computer Engineering, \\McMaster
University, Hamilton, Ontario, Canada.\\
Email: jkzhang@mail.ece.mcmaster.ca\\
}
\and
\authorblockN{Chau Yuen}
\authorblockA{Institute for Infocomm Research, Room 03-07\\
 21 Heng Mui Keng Terrace, Singapore 119613.\\
Email:  cyuen@i2r.a-star.edu.sg} \and
 }
%

\maketitle

\begin{abstract}
In this paper, we develop two kinds of novel closed-form
decompositions on phase shift keying (PSK) constellations by
exploiting linear congruence equation theory: the one for
factorizing a $pq$-PSK constellation into a product of a $p$-PSK
constellation and a $q$-PSK constellation, and the other for
decomposing a specific complex number into a difference of a
$p$-PSK constellation and a $q$-PSK constellation. With this, we
propose a simple signal design technique to blindly and uniquely
identify frequency selective channels with zero-padded block
transmission under noise-free environments by only using the first
two block received signal vectors. Furthermore, a closed-form
solution to determine the transmitted signals and the channel
coefficients is obtained. In the Gaussian noise and Rayleigh
fading environment, we prove that the newly proposed signaling
scheme enables non-coherent full diversity for the Generalized
Likelihood Ratio Test (GLRT) receiver.
\end{abstract}

\section{Introduction}

In this paper, we consider wireless communication systems with a
single transmitting antenna and a single receiving antenna which
transmit data over a frequency-selective fading channel. The
systems which we consider mitigate the intersymbol interference
generated by the channel by transmitting the data stream in
consecutive equal-size blocks, which are subsequently processed at
the receiver on a block-by-block basis, see,
e.g.,~\cite{giannakis99, wang00, jkz-iee05}. In order to remove
interblock interference, some redundancy is added to each block
before transmission. There are several ways to add
redundancy~(e.g.,\cite{giannakis99,wang00}), but in this paper we
will focus on block-by-block communication systems with
zero-padding redundancy; e.g.,~\cite{wang00, giannakis99,
jkz-iee05}.

When the receiver possesses perfect knowledge of the channel and
employs maximum likelihood (ML) detection, it was
shown~\cite{jkz-iee05, Wang_Ma_GG_2002} that such a system not
only enables full diversity but also provide the maximum coding
gain. Unfortunately, perfect channel state information at the
receiver, in practice, is not easily attainable. If the coherent
time is sufficiently long, then, the transmitter can send training
signals that allow the receiver to estimate the channel
coefficients accurately. For mobile wireless communications,
however, the fading coefficients may change so rapidly that the
coherent time may be too short to allow reliable estimation of the
coefficients, especially in a system with a large number of
antennas. Therefore, the time spent on sending training signals
cannot be ignored because of the necessity of sending more
training signals for the accurate estimation of the
channel~\cite{hassibi02, zheng02}.

To avoid having to transmit training signals, considerable
research efforts have been directed to develop techniques of
``blind channel estimation"~\cite{ALLE98, Tong98} recently. These
techniques identify and estimate the transmission channel using
only the received (perhaps noisy) signals at the receiver. The
essence of these algorithms is to exploit the structure of the
channel and/or the property of transmitted signals. The subspace
method is one such method exploiting the channel structure and the
second order statistics of input signals~\cite{Tong98, scaglione99
}. In digital communication applications, the input signals have
finite alphabet property such as the constant modulus for PSK
modulation or integers for QAM, which can be further exploited to
estimate the channel with shorter coherent time than the subspace
method by numerically solving some optimization
problems~\cite{Ding00}. Thus far, all currently available blind
methods without either training or pilots for the frequency
selection channel estimation incur scale ambiguity and as a
consequence, cannot \textit{uniquely} identify the channel
coefficients. In addition, wireless communication applications
demand the accurate estimate of the channel as well as of the
signal. The resulting scale ambiguity from the channel estimation
will result in significant error probability for the signal
estimation even in a noise-free environment.

To resolve this issue, we propose a novel signaling and
transmitting technique for the frequency selective channel with
zero-padding block transmission, in which neither the transmitter
or the receiver knows the channel state information. Our main
contributions in this paper are as follows:
\begin{enumerate}
 \item A novel signal design technique using a pair of coprime
$p$-PSK and $q$-PSK constellations for the first two block
transmissions is proposed to blindly and uniquely identify
frequency selective channels with zero-padded block transmission
by only processing the first two block received signals.
Furthermore, a closed-form solution to determine the transmitted
signals and the channel coefficients is obtained by utilizing the
linear congruence equation theory.
 \item In the Gaussian noise and Rayleigh fading environment, we
 prove that the newly proposed scheme enables full diversity for
 the GLRT receiver.
\end{enumerate}
Here, we should point out that the similar hybrid signaling
schemes~\cite{Zhou05, cui07, zhou07, ma07, ken08} were used to
eliminate the ambiguity of the blind orthogonal space-time block
codes.

{\bf Notation}: Column vectors and matrices are boldface lowercase
and uppercase letters, respectively; the matrix transpose, the
complex conjugate, the Hermitian are denoted by
$(\cdot)^T,(\cdot)^*, (\cdot)^H$, respectively; $\mathbf{I}_N$
denotes the $N\times N$ identity matrix; $\mathrm{gcd}(m,n)$
denotes the greatest common divisor of positive integers $m$ and
$n$. Particularly when $\gcd(m,n) = 1$, we say that $m$ and $n$
are coprime integers; $\varphi(n)$ denotes the Euler function,
i.e., the number of all positive integers that do not exceed $n$
and are prime to $n$. The $(i, j)$th element of matrix
$\mathbf{A}$ is denoted by $a_{ij}$.

\section{Channel model and blind signal designs}\label{sec:model} If the channel is assumed to be of length at
most $L$ (i.e., if $L$ is an upper bound on the delay spread),
then, the block transmission system with zero-padding operate as
follows: First, $L-1$ zeros are append to ${\mathbf x}$ to form
${\mathbf x}'$ which is of length $P=K+L-1$. The elements of
${\mathbf x}'$ are serially transmitted through the channel. The
channel impulse response is denoted by ${\mathbf h}=[h_0, h_1,
\cdots, h_{L-1}]^T$. The length $P$ received signal vector
${\mathbf r}$ can be written as
\begin{equation}\label{model1}
{\mathbf r}={\mathbf H}{\mathbf s}+\boldsymbol{\xi},
\end{equation}
where ${\mathbf r}$ is a $P\times 1$ received signal vector,
${\mathbf s}$ is a $K\times 1$ transmitted signal vector,
$\boldsymbol{\xi}$ denotes the $P\times 1$ vector of noise samples
at the receiver, and ${\mathbf H}$ denotes the $P\times K$
Toeplitz matrix~\cite{giannakis99,wang00, jkz-iee05}
\begin{small}
\begin{displaymath}
{\mathbf H}=\left( \begin{array}{cccc}
h_0 & 0 & \ldots & 0\\
h_1 & h_0 & \ldots & 0\\
\vdots & h_1 & \ddots & \vdots\\
h_{L-1} & \ddots & \ddots & h_0\\
0 & \ddots & \ddots & h_1\\
\vdots & \ddots & \ddots & \vdots\\
0 & \ddots & 0 & h_{L-1}
\end{array}\right)_{P\times K}.
\end{displaymath}
\end{small}
For blind channel estimation, it is convenient to rewrite the
channel model~(\ref{model1}) as
\begin{equation}\label{model2}
{\mathbf r}={\mathcal T}({\mathbf s}){\mathbf h}+\boldsymbol{\xi},
\end{equation}
where we have used the following fact ${\mathbf H}{\mathbf
s}={\mathcal T}({\mathbf s}){\mathbf h}$
with ${\mathcal T}({\mathbf s})$ defined by
\begin{small}
\begin{displaymath}
{\mathcal T}({\mathbf s})=\left( \begin{array}{cccc}
s_1 & 0 & \ldots & 0\\
s_2 & s_1 & \ldots & 0\\
\vdots & s_2 & \ddots & \vdots\\
s_K & \ddots & \ddots & s_1\\
0 & \ddots & \ddots & s_2\\
\vdots & \ddots & \ddots & \vdots\\
0 & \ddots & 0 & s_K
\end{array}\right)_{P\times L}.
\end{displaymath}
\end{small}
Now we assume that during a $T$ block transmission period, the
channel coefficients keep constant and after that, will randomly
change. Our novel blind modulation scheme is described as follows:
During the first block transmission, each symbol of a transmitted
signal vector ${\mathbf s}={\mathbf x}$ is chosen from $p$-PSK
constellation ${\mathcal X}$; i.e.,
${\mathbf r}_1={\mathcal T}({\mathbf x}){\mathbf
h}+\boldsymbol{\xi}_1,\qquad {\mathbf x}\in{\mathcal X}^K$.
During the second block transmission, each symbol
of a transmitted signal vector ${\mathbf s}={\mathbf y}$ is chosen
from $q$-PSK constellation ${\mathcal Y}$; i.e.,
${\mathbf r}_2={\mathcal T}({\mathbf y}){\mathbf
h}+\boldsymbol{\xi}_2,\qquad {\mathbf y}\in{\mathcal Y}^K$,
where $p$ and $q$ are coprime. Then, during the $i$th block
transmission for $3\le i\le T$, each symbol of a transmitted
signal vector ${\mathbf s}={\mathbf z}_i$ can be chosen from any
constellation ${\mathcal Z}$; i.e.,
${\mathbf r}_i={\mathcal T}({\mathbf z}_i){\mathbf
h}+\boldsymbol{\xi}_i,\qquad {\mathbf z}_i\in{\mathcal Z}^K$.
Collecting all the $T$ block received signals, we have
\begin{subequations}\label{blind modulation}
\begin{eqnarray}
{\mathbf z}={\mathbf S}{\mathbf h}+{\boldsymbol \eta},
\end{eqnarray}
where ${\boldsymbol \eta}=(\boldsymbol{\xi}^T_1,
\boldsymbol{\xi}^T_2, \cdots, \boldsymbol{\xi}^T_T)^T$ and
\begin{eqnarray}
{\mathbf S}=\big({\mathcal T}^T({\mathbf x}), {\mathcal
T}^T({\mathbf y}), {\mathcal T}^T({\mathbf z}_3), \cdots,
{\mathcal T}^T({\mathbf z}_T)\big)^T
\end{eqnarray}
for ${\mathbf x}\in {\mathcal X}^K, {\mathbf y}\in {\mathcal Y}^K,
{\mathbf z}_i\in{\mathcal Z}^K$.
\end{subequations}
Throughout this paper, we make the following assumptions:
\begin{enumerate}
 \item The channel coefficients $h_\ell$ for $\ell =0, 1, \cdots, L-1$ are samples of independent circularly symmetric zero-mean complex white
    Gaussian random variables with unit variances and remain constant for the first $PT$ ($T\ge 2$) time slots,
after which they change to new independent values that are fixed
for next $PT$ time slots, and so on.
 \item
The elements $\boldsymbol\eta$ is circularly
    symmetric zero-mean complex Gaussian samples with covariance matrix $\sigma^2{\mathbf
    I}_{PT}$;
 \item During $PT$ observable time slots,
$T$ consecutive blocks ${\mathbf z}_i$ are transmitted with each
entry $z_{ik}$ for $i=3, 4, \cdots, T$ and $k=1, 2, \cdots, P$
being independently and equally likely chosen from the
constellation ${\mathcal Z}$, while components $x_k$ and $y_k$ for
$k=1, 2, \cdots, P$ in the previous two blocks are independently
and equally likely chosen from the respective $p$-PSK and $q$-PSK
constellations ${\mathcal X}$ and ${\mathcal Y}$, where $p$ and
$q$ are coprime.
 \item Channel state
information is not available at either the transmitter or the
receiver.
\end{enumerate}

Our goal in this paper is to prove that our designed signaling
scheme (\ref{blind modulation})
\begin{enumerate}
 \item enables the unique identification of the channel and the
 transmitted signals for any given nonzero received signal vector ${\mathbf r}$ in a noise-free case and
 \item provides full diversity for the GLRT receiver in the Gaussian noise
 and Rayleigh fading environment.
\end{enumerate}

\section{Blind unique identification and full diversity}
In this section, we first develop some decomposition properties on
a pair of PSK constellations and then, prove that our blind
modulation scheme proposed in the previous section enables the
unique identification of the channel coefficients and the
transmitted signals in a noise-free case as well as full diversity
for the GLRT receiver in a noise environment.
\subsection{Decompositions of PSK
Constellations}
\begin{proposition} \label{pro:factor}
Let two positive integers $p$ and $q$ be coprime. Then, for any
integer $k$ with $0 \leq k < pq$, there exists a pair of $x$ and
$y$ such that
\begin{equation}\label{eq:spsq-1}
    x y= \exp\left(j\frac{2\pi k}{pq}\right).
\end{equation}
Furthermore, $s_{p}$ and $s_{q}$ can be uniquely determined by
\begin{subequations}\label{eq:spq}
    \begin{eqnarray}
        x &=& \exp\left(j\frac{2\pi k\,
        q^{\varphi(p)-1}}{p}\right)\\
        y &=& \exp\left(j\frac{2\pi k\,
        p^{\varphi(q)-1}}{q}\right).
    \end{eqnarray}
\end{subequations}
~\hfill\QED
\end{proposition}
Proposition~\ref{pro:factor} tells us that any $pq$-PSK symbol can
be uniquely factored into the product of a pair of the coprime
$p$-PSK symbol and $q$-PSK symbol. This factorization was first
discovered by Zhou, Zhang and Wong~\cite{Zhou05, zhou07}. Now,
Proposition~\ref{pro:factor} significantly simplifies the original
representation. The following property gives a necessary and
sufficient condition for a complex number to be able to be
decomposed into a difference of a pair of  the coprime $p$-PSK
symbol and $q$-PSK symbol.
\begin{proposition}\label{pro:unit}
Let $w\ne 0$ be a given non-zero complex number and
$w=|w|e^{j\theta}$. Then, there exists a pair of $x\in {\mathcal
X}$ and $y\in{\mathcal Y}$ satisfying equation
\begin{eqnarray}\label{eq:unit}
x-y=w
\end{eqnarray}
if and only if there exist three integers $m, n$ and $k$ such that
\begin{subequations}\label{condt}
\begin{eqnarray}
\theta&=&\frac{\pi(p n+q m)}{pq}+k\pi+\frac{\pi}{2}\\
|w|&=&(-1)^{k+1} \sin\left(\frac{\pi(p n-q m)}{pq}\right).
\end{eqnarray}
\end{subequations}
Furthermore, under the condition~(\ref{condt}), if $p$ and $q$ are
coprime, then, equation~(\ref{eq:unit}) has the unique solution
that can be explicitly determined as follows:
\begin{enumerate}
 \item If $w=0$, then, $x=y=1$.
 \item If $w\ne 0$, then, $x$ and $y$ are given by
\begin{subequations}\label{eq:spq}
    \begin{eqnarray}
        x &=& \exp\left(j\frac{2\pi \ell q^{\varphi(p)-1}}{p}\right)\\
        y &=& \exp\left(j\frac{2\pi \ell p^{\varphi(q)-1}}{q}\right)
    \end{eqnarray}
\end{subequations}
with integer $\ell
=pq\left(\frac{\theta}{\pi}-\frac{1}{2}\right)$.
\end{enumerate}
~\hfill\QED
\end{proposition}
The proofs of Propositions~\ref{pro:factor} and~\ref{pro:unit} are
omitted because of space limitation.

\subsection{Blind unique identification of the channel}
Our main purpose in this subsection is to prove that the blind
signaling scheme~(\ref{blind modulation}) is capable of uniquely
identifying the channel coefficients and the transmitted symbols.
To do this, Let ${\mathbf u}$ and ${\mathbf v}$ be two consecutive
block received signal vectors in the first two block transmission
from the channel model~(\ref{model2}) in a noise-free environment;
i.e.,
\begin{eqnarray}
{\mathbf u}&=&{\mathcal T}({\mathbf x}){\mathbf
h}\qquad {\mathbf x}\in{\mathcal X}^K,\label{eq:blind-x}\\
{\mathbf v}&=&{\mathcal T}({\mathbf y}){\mathbf h}\qquad {\mathbf
y}\in{\mathcal Y}^K.\label{eq:blind-y}
\end{eqnarray}
Now, we formally state the first result.
\begin{theorem}(Unique Identification)\label{th:identi}
Let ${\mathbf u}=(u_1, u_2, \cdots, u_P)^T$ and ${\mathbf v}=(v_1,
v_2, \cdots, v_P)^T$ be the first two consecutive block nonzero
received signal vectors given by~(\ref{eq:blind-x})
and~(\ref{eq:blind-y}), respectively, and $p$ and $q$ be co-prime.
If $r$ denotes the maximum integers such that $u_1 v_1=u_2
v_2=\cdots=u_r v_r=0$, then, $h_0=h_1=\cdots=h_{r-1}=0$. In
addition, the other remaining channel coefficients $h_{r},
h_{r+1},\cdots, h_{L-1}$ and all the transmitted symbols in
${\mathbf x}$ and ${\mathbf y}$ can be uniquely determined as
follows.
\begin{enumerate}
 \item Let $w_1$ be defined by $w_1=\frac{u_{r+1}}{v_{r+1}}=|w_1|e^{j\theta_1}$ and
 $\ell_1=\frac{pq\theta_1}{2\pi}$. Then,
 we have
 \begin{subequations}\label{solu:x1y1}
    \begin{eqnarray}
        x_1 &=& \exp\left(j\frac{2\pi \ell_1 q^{\varphi(p)-1}}{p}\right)\label{solu:x1}\\
        y_1 &=& \exp\left(-j\frac{2\pi \ell_1
        p^{\varphi(q)-1}}{q}\right)\label{solu:y1}\\
        h_r&=& x_1^* u_{r+1}\label{solu:hr}.
    \end{eqnarray}
\end{subequations}
 \item For $1<m\le L-r$, let $w_m$ be defined by
\begin{eqnarray}\label{def:w}
 w_m &=&\frac{x_1^*\left(u_{r+m}-\sum_{i=1}^{m-2}
 h_{r+i}x_{m-i}\right)}{h_r}\nonumber\\
&& -\frac{y_1^*\left(v_\ell-\sum_{i=1}^{m-2} h_{r+i}
x_{m-i}\right)}{h_r}.
\end{eqnarray}
\begin{enumerate}
 \item If $w_m=0$, then, we have
 \begin{small}
\begin{subequations}
\begin{eqnarray}
x_m &=& x_1\label{solu:xwo}\\
y_m &=& y_1\label{solu:ywo}\\
h_{r+m-1} &=& x_1^*\big(u_{r+m}-\sum_{i=1}^{m-2} h_{r+i}
x_{m-i}\big)\label{solu:hwo}.
\end{eqnarray}
\end{subequations}
\end{small}
 \item If $w_m\ne 0$, let $w_m=|w_m| e^{j\theta_m}$ and $\ell_m=pq\left(\frac{\theta_m}{\pi}-\frac{1}{2}\right)$.
 Then, we have
 \begin{small}
 \begin{subequations}
\begin{eqnarray}
x_m &=& x_1 \exp\big(j\frac{2\pi \ell_m q^{\varphi(p)-1}}{p}\big)\label{solu:xm} \\
y_m &=& y_1 \exp\big(j\frac{2\pi \ell_m
        p^{\varphi(q)-1}}{q}\big)\label{solu:ym}\\
h_{r+m-1} &=& x_1^*\big(u_{r+m}-\sum_{i=1}^{m-2} h_{r+i}
x_{m-i}\big)\label{solu:hm}.
\end{eqnarray}
\end{subequations}
\end{small}
\end{enumerate}
\item For $L-r+1\le m\le K$, we have
\begin{small}
\begin{subequations}\label{solu:xmym}
\begin{eqnarray}
x_m &=& \frac{u_{m+r}-\sum_{i=1}^{L-r-1} h_{L-i}\,
x_{m-L+r+i}}{h_r}\\
y_m &=& \frac{v_{m+r}-\sum_{i=1}^{L-r-1} h_{L-i}\,
y_{m-L+r+i}}{h_r}.
\end{eqnarray}
\end{subequations}
\end{small}
\end{enumerate}
 ~\hfill\QED
\end{theorem}
\textit{Proof}: Basically, the proof of Theorem~\ref{th:identi}
captures the following four steps.

{\bf Step~1}.  First, we consider the first received signals in
each block. In this case, we have
\begin{subequations}\label{eq:th-prof-case1}
\begin{eqnarray}
h_0 x_1=u_1,\\
h_0 y_1=v_1.
\end{eqnarray}
\end{subequations}
Therefore, if either $u_1$ or $v_1$ is zero, then, $h_0$ is zero.
Similarly, we can obtain $h_1=h_2=\cdots, h_{r-1}=0$ if
$u_1v_1=u_2v_2=\cdots=u_rv_r=0$.

{\bf Step~2}. We continue to proceed the $(r+1)$-th received
signals for each block. In this case, we have received
\begin{subequations}\label{eq:th-prof-case2}
\begin{eqnarray}
h_r x_1&=&u_{r+1},\\
h_r y_1&=& v_{r+1}.
\end{eqnarray}
\end{subequations}
Since $u_{r+1}v_{r+1}\ne 0$, eliminating $h_1$
from~(\ref{eq:th-prof-case2}) results in
\begin{eqnarray}
x_1y_1^*=\frac{u_{r+1}}{v_{r+1}}=w_1.
\end{eqnarray}
Now, by Proposition~\ref{pro:factor}, $x_1$ and $y_1$ can be
uniquely determined by~(\ref{solu:x1}) and~(\ref{solu:y1}),
respectively, and thus, $h_r$ is uniquely determined
by~(\ref{solu:hr}).

{\bf Step 3}. Let us consider the $(r+2)$-th received signals for
each block:
\begin{subequations}\label{eq:th-prof-case3}
\begin{eqnarray}
h_{r+1}x_1+h_rx_2&=&u_{r+2},\\
h_{r+1}y_1+h_ry_2&=&v_{r+2}.
\end{eqnarray}
\end{subequations}
Eliminating $h_{r+1}$ from~(\ref{eq:th-prof-case3}) yields
\begin{eqnarray}
x_1^*x_2-y_1^* y_2=\frac{u_{r+2} x_1^*-v_{r+2} y_1^*}{h_r}=w_2
\end{eqnarray}
Since $x_i\in {\mathcal X}$ and $y_i\in {\mathcal Y}$ for $i=1,
2$, we have $x_1^* x_2\in {\mathcal X}$ and $y_1^* y_2\in{\mathcal
Y}$ too. Now, by Proposition~\ref{pro:unit}, $x_2$ and $y_2$ can
be uniquely determined by~(\ref{solu:xwo}) or~(\ref{solu:xm})
and~(\ref{solu:ywo}) or~(\ref{solu:ym}), respectively, and thus,
$h_{r+1}$ is uniquely determined by~(\ref{solu:hwo})
or~(\ref{solu:hm}) with $m=2$.

In general, we proceed to consider determining the $(r+m)$-th
channel coefficient for $2<m\le L-r-1$. In this case, we have
received
\begin{subequations}
\begin{eqnarray}
h_{r+m-1}x_1+h_{r+m-2}x_2+\cdots+h_r x_m&=&u_{r+m}\\
h_{r+m-1}y_1+h_{r+m-2}y_2+\cdots+h_r y_m&=&v_{r+m}
\end{eqnarray}
\end{subequations}
This is equivalent to
\begin{subequations}\label{eq:th-prof-generalcase}
\begin{eqnarray}
h_{r+m-1}x_1+h_r x_m&=&u_{r+m}-\sum_{i=1}^{m-2} h_{r+i} x_{m-i}\\
h_{r+m-1}y_1+h_r y_m&=&v_{r+m}-\sum_{i=1}^{m-2} h_{r+i} x_{m-i}
\end{eqnarray}
\end{subequations}
Eliminating $h_{r+m-1}$ from~(\ref{eq:th-prof-generalcase}) yields
\begin{eqnarray}
&&x_1^* x_m-y_1^* y_m=\frac{x_1^*\left(u_{r+m}-\sum_{i=1}^{m-2}
h_{r+i}
x_{m-i}\right)}{h_r}\nonumber\\
&&-\frac{y_1^*\left(v_{r+m}-\sum_{i=1}^{m-2} h_{r+i}
x_{m-i}\right)}{h_r}=w_m.
\end{eqnarray}
Now, by Proposition~\ref{pro:unit}, $x_m$ and $y_m$ can be
uniquely determined by~(\ref{solu:xwo}) or~(\ref{solu:xm})
and~(\ref{solu:ywo}) or~(\ref{solu:ym}), respectively, and thus,
$h_{r+m-1}$ is uniquely determined by~(\ref{solu:hwo})
or~(\ref{solu:hm}).

 {\bf Step 4}. $L-r-1<m\le K$. In this case, since the channel
 coefficients have been determined by the previous three steps,
 we can determine the other remaining transmitted
 signals from the remaining received signals of each block. In
 this case, we have
 \begin{small}
\begin{subequations}
\begin{eqnarray}
h_{L-1}x_{m-L+r+1}+h_{L-2}x_{m-L+r+2}+\cdots+h_r x_m=u_{r+m}\\
h_{L-1}y_{m-L+r+1}+h_{L-2}y_{m-L+r+2}+\cdots+h_r y_m=v_{r+m}
\end{eqnarray}
\end{subequations}
\end{small}
From this we can obtain~(\ref{solu:xmym}). This completes the
proof of Theorem~\ref{th:identi}.~\hfill$\Box$

We would like to make the following observations on
Theorem~\ref{th:identi}.
\begin{enumerate}
 \item Theorem~\ref{th:identi} not only tells us that the channel
coefficients can be uniquely identified by {\em only} transmitting
two block signals with each symbol selected from two co-prime PSK
constellations, but also provides simple and closed-form solutions
to both the channel coefficients and the transmitted symbols. The
traditional blind method~\cite{scaglione99} based on the second
order statistics requires a lot of data blocks. Even so, it still
cannot uniquely identify the channel coefficients as well as the
transmitted signals.
 \item If we set $p=2^m$ and $q=2^m+1$ with $m$ being a positive
 integer, then, it is clear that $p$
 and $q$ are coprime. Therefore, Theorems~\ref{th:identi}
 holds for such a pair of $p$ and $q$. In
 addition, if we want the original symbol sets ${\mathcal X}$
 and ${\mathcal Y}$ to contain the same integer bits, we can
 delete the only one common element 1 from ${\mathcal Y}$; i.e.,
 $\overline{\mathcal Y}={\mathcal Y}-\{1\}$. Thus, there are
 totally
 $2^m$ elements in the remaining set $\overline{\mathcal Y}$ and Theorems~\ref{th:identi}
 still holds for such a pair of constellations
 ${\mathcal X}$ and $\overline{\mathcal Y}$.
 \item By Theorem~\ref{th:identi}, if we let $m_0$ denote the maximum positive integer such that
$u_{r+m}-\sum_{i=1}^{m-2} h_{r+i}x_{m-i}=0$ for $m_0<m\le L-r$ but
$u_{r+m_0}-\sum_{i=1}^{m_0-2} h_{r+i}x_{m_0-i}=0$, then, the
length of the channel is actually equal to $m_0-r+1$. Therefore,
our blind modulation scheme
 enables the receiver to exactly determine the length of the channel by only utilizing the first
 two block received signals.
\end{enumerate}

\begin{figure*}[!b]
\hrulefill
\begin{small}
\begin{eqnarray}\label{flat matrix}
&&\left|%
\begin{array}{cc}
  {\mathcal T}({\mathbf x})[1:K] & {\mathcal T}(\widetilde{\mathbf x})[1:K] \\
  {\mathcal T}({\mathbf z}_i)[k:K+k-1] & {\mathcal T}(\widetilde{\mathbf z}_i)[k:K+k-1] \\
\end{array}%
\right|\nonumber
=\left|%
\begin{array}{cc}
  {\mathcal T}({\mathbf x})[1:K] & {\mathcal T}(\widetilde{\mathbf x})[1:K]-{\mathcal T}({\mathbf x})[1:K]\\
  {\mathcal T}({\mathbf z}_i)[k:K+k-1] & {\mathcal T}(\widetilde{\mathbf z}_i)[k:K+k-1]-{\mathcal T}({\mathbf z}_i)[k:K+k-1] \\
\end{array}%
\right|\nonumber\\
&&=\left|%
\begin{array}{cc}
  {\mathcal T}({\mathbf x})[1:K] & {\mathbf 0} \\
  {\mathcal T}({\mathbf z}_i)[k:K+k-1] & {\mathcal T}(\widetilde{\mathbf z}_i)[k:K+k-1]-{\mathcal T}({\mathbf z}_i)[k:K+k-1] \\
\end{array}%
\right|=x_1^K(z_{ik}-\widetilde z_{ik})^K\ne 0.
\end{eqnarray}
\end{small}
\end{figure*}

\subsection{Full diversity}
The GLRT requires neither the knowledge of the fading and noise
statistics, nor the knowledge of their
realizations\cite{narayan98}. 
The criterion can be simply stated as
$\hat{{\mathbf S}}=\arg\max_{{\mathbf S}}\{{\mathbf z}^H{\mathbf
S}\left({\mathbf S}^H{\mathbf S}\right)^{-1}{\mathbf S}^H{\mathbf
z}\}$
In fact, the GLRT projects the received signal ${\mathbf z}$ on
the different subspaces spanned by ${\mathbf S}$ and then
calculate the energies of all the projections and choose the
projection that maximizes the energy. Now, in order to examine
full diversity, for any pair of distinct codewords ${\mathbf S}$
and $\widetilde{\mathbf S}$, let $\left(
  \begin{array}{c}
    {\mathbf S}^H \\
    \widetilde{\mathbf S}^H \\
  \end{array}
\right)\left({\mathbf S}, \widetilde{\mathbf S}\right)={\mathbf
A}.$
Brehler and Varanasi~\cite{brehler01} proved the following lemma.
\begin{lemma}\label{lem:brehler}
If matrices ${\mathbf A}$ have full column rank for all pair of
distinct codewords ${\mathbf S}$ and $\widetilde{\mathbf S}$,
then, the code-book provides full diversity for the GLRT
receiver.~\hfill\QED
\end{lemma}

Now, we are in position to formally state the second our main
result.
\begin{theorem}\label{th:fd} The blind modulation designed
in Section~\ref{sec:model} with $p$ and $q$ being coprime enables
full diversity for the GLRT receiver.~\hfill\QED
\end{theorem}
\textit{Proof}: By Lemma~\ref{lem:brehler}, we only need to prove
that $({\mathbf S}, \widetilde{\mathbf S})$ has full column rank
for any pair of distinct signal matrices ${\mathbf S}$ and
$\widetilde{\mathbf S}$. Otherwise, if there existed a pair of
distinct codeword matrices ${\mathbf S}$ and $\widetilde{\mathbf
S}$ for which the matrix $({\mathbf S},\widetilde{\mathbf S})$
does not have full column rank, then, the linear equations with
respect to variables ${\mathbf h}$ and $-\widetilde{\mathbf h}$,
$\left(%
\begin{array}{cc}
  {\mathbf S} & \widetilde{\mathbf S}
\end{array}%
\right)\left(%
\begin{array}{c}
  {\mathbf h} \\
  -\widetilde{\mathbf h} \\
\end{array}%
\right)={\mathbf 0}$,
would have a nonzero solution ${\mathbf h}_0$ and
$\widetilde{\mathbf h}_0$. Let ${\mathbf r}_0={\mathbf S}{\mathbf
h}_0$. Then, we would also have ${\mathbf r}_0=\widetilde{\mathbf
S}\widetilde{\mathbf h}_0$. In other words, for a given nonzero
received signal ${\mathbf r}_0$, equation ${\mathbf r}_0={\mathbf
S}{\mathbf h}$ has two distinct pair of solutions. By
Theorem~\ref{th:identi}, we have that ${\mathcal T}({\mathbf
x})={\mathcal T}(\widetilde{\mathbf x})$ and ${\mathcal
T}({\mathbf y})={\mathcal T}(\widetilde{\mathbf y})$. Since
${\mathbf S}\ne\widetilde{\mathbf S}$, there is a pair of distinct
signal sub-matrices in ${\mathbf S}$ and $\widetilde{\mathbf S}$,
${\mathcal T}({\mathbf z}_i)$ and ${\mathcal T}(\widetilde{\mathbf
z}_i)$ for some $3\le i\le L$. That being said, ${\mathbf z}_i\ne
\widetilde{\mathbf z}_i$. If we let ${\mathbf z}_i=(z_{i1},
z_{i2}, \cdots, z_{iK})^T$ and $\widetilde{\mathbf
z}_i=(\widetilde z_{i1}, \widetilde z_{i2}, \cdots, \widetilde
z_{iK})^T$, then, there exists a positive integer $k$ such that
$z_{ik}\ne \widetilde z_{ik}$ but $z_{i\ell}=\widetilde z_{i\ell}$
for $\ell=1, 2, \cdots, k-1$. For notional simplicity, we use
${\mathbf B}[M:N]$ to denote the sub-matrix of a matrix ${\mathbf
B}$ consisting of all the columns and the rows from $M$ to $N$.
Then, we have the result~\eqref{flat matrix}, which is shown in
the bottom of this page,
where we have used the fact that ${\mathcal T}(\widetilde{\mathbf
z}_i)[k:K+k-1]-{\mathcal T}({\mathbf z}_i)[k:K+k-1]$ is actually a
$K\times K$ lower triangular matrix with the diagonal entries
being all equal to $z_{ik}-\widetilde z_{ik}$.
Therefore, the sub-matrix of $({\mathbf S}, \widetilde{\mathbf S})$, $\left(%
\begin{array}{cc}
  {\mathcal T}({\mathbf x}_i)[1:K] & {\mathcal T}(\widetilde{\mathbf x})[1:K] \\
  {\mathcal T}({\mathbf z}_i)[k:K+k-1] & {\mathcal T}(\widetilde{\mathbf z}_i)[k:K+k-1] \\
\end{array}%
\right)$ is a $2K\times 2K$ invertible matrix and hence,
$({\mathbf S}, \widetilde{\mathbf S})$ has full column rank, which
contradicts with the previous assumption. This completes the proof
of Theorem~\ref{th:fd}.~\hfill$\Box$

So far, we have shown
 that our blind modulation scheme not only enables the unique
 identification of the channel coefficients in the noise-free case but also full
 diversity in the noise environment. However, the
traditional blind method~\cite{scaglione99} based on the second
order statistics can provide neither the unique identification of
the channel coefficients nor estimation of the transmitted signals
with full diversity reliability. Similar to the Comment~2) on
Theorem~\ref{th:identi}, our Theorem~\ref{th:fd} is also true
 for both a particular pair of
 $p=2^m$ and $q=2^m+1$ and the derived pair of constellations.

\section{Conclusion}\label{sec8}

In this paper, we proposed a novel blind modulation technique to
uniquely identify frequency selective channels with zero-padded
block transmission under noise-free environments by only
processing the first two block received signal vectors.
Furthermore, a closed-form solution to determine the transmitted
signals and the channel coefficients was derived by using linear
congruence equation theory. In the Gaussian noise and Rayleigh
fading environment, we proved that our new scheme enables full
diversity for the GLRT receiver. 

\bibliographystyle{ieeetr}
\bibliography{tzzt}
\end{document}